\documentclass[twoside]{article}
\usepackage{multirow} 
\usepackage{arxiv}

\usepackage{graphicx}       
\usepackage{tabularx}
\usepackage{booktabs}
\usepackage{geometry}
\usepackage{amsmath}

\title{COVID-19 Pneumonia Diagnosis Using Medical Images: Deep Learning-Based Transfer Learning Approach}

\author{
  Anjali Dharmik \\
  Senior Data Scientist \\
  HP, Inc \\
}

\begin{document}
\maketitle

\begin{abstract}
\subsection{Background}
SARS-CoV-2, the causative agent of COVID-19, remains a global health concern due to its high transmissibility and evolving variants \cite{ref1}. Although vaccination efforts and therapeutic advancements have mitigated disease severity, emerging mutations continue to challenge diagnostics and containment strategies. As of mid-February 2025, global test positivity has risen to 11\% \cite{ref1}, marking the highest level in over six months despite widespread immunization efforts \cite{ref2}. Newer variants demonstrate enhanced host cell binding, increasing both infectivity and diagnostic complexity.

\subsection{Objective}
This study evaluates the effectiveness of deep transfer learning in delivering rapid, accurate, and mutation-resilient COVID-19 diagnosis from medical imaging, with a focus on scalability and accessibility.

\subsection{Method}
We developed an automated detection system using state-of-the-art CNNs, including VGG16, ResNet50, ConvNetXtTiny, MobileNet, NASNetMobile, and DenseNet121 among others, to detect COVID-19 from chest X-ray and CT images.

\subsection{Result}
Among all the models evaluated, DenseNet121 emerged as the best-performing architecture for COVID-19 diagnosis using CT and X-ray images. It achieved an impressive accuracy of 98\%, with 96.9\% precision, 98.9\% recall, 97.9\% F1-score and 99.8\% AUC score, indicating a high degree of consistency and reliability in both detecting positive and negative cases. The confusion matrix showed minimal false positives and false negatives, underscoring the model's robustness in real-world diagnostic scenarios. Given its performance, DenseNet121 is a strong candidate for deployment in clinical settings and serves as a benchmark for future improvements in AI-assisted diagnostic tools.

\subsection{Conclusion}
These results underscore the potential of AI-powered diagnostics in supporting early detection and global pandemic response. With careful optimization, deep learning models can address critical gaps in testing, particularly in settings constrained by limited resources or emerging variants.

\end{abstract}

\keywords{{Computer Vision \and COVID-19 Pneumonia Diagnosis \and Deep Learning \and Transfer Learning \and Medical Imaging Analysis}}

\section{Introduction}

\subsection{Background}
SARS-CoV-2, the virus responsible for COVID-19, first emerged on 31 December 2019 in Wuhan City, Hubei Province, China \cite{ref6}. It is a highly transmissible respiratory pathogen capable of causing severe illness or death across all age groups \cite{ref7}. Since its initial outbreak, substantial progress has been made in managing the virus through vaccination, antiviral therapies, and AI-powered diagnostic technologies.

Despite these advances, SARS-CoV-2 continues to pose a global health challenge, especially for immunocompromised individuals and those with underlying conditions. One of the most persistent obstacles is the virus's ability to mutate rapidly. To date, more than 26 genetically distinct variants have been identified, many of which exhibit increased transmissibility and immune evasion due to mutations that enhance their binding affinity to host cells \cite{ref2}.

By 20 August 2023, the pandemic had resulted in over 769 million confirmed cases and more than 6.9 million deaths worldwide \cite{ref9}. Early in the pandemic, the WHO declared COVID-19 a PHEIC on 30 January 2020 \cite{ref10}.

More recently, SARS-CoV-2 has shown a global resurgence. As of 11 May 2025, surveillance data from the GISRS indicated that the global test positivity rate reached 11\%, up significantly from 2\% in February 2025 \cite{ref1}. This current wave, comparable to the July 2024 peak of 12\%, is largely driven by cases in the Eastern Mediterranean, South-East Asia, and Western Pacific Regions \cite{ref1}.

A key driver of this resurgence is the emergence of the recombinant XEC variant, first detected in Germany in June 2024 \cite{ref3}. Derived from two Omicron subvariants KS.1.1 and KP.3.3, XEC rapidly spread worldwide and by December 2024 accounted for nearly 45\% of cases in the United States \cite{ref2, ref3, ref4, ref5}. Its global dominance underscores the critical importance of continued genomic surveillance and adaptive diagnostic strategies.

In February 2025, the WHO categorized circulating variants as follows:
\begin{itemize}
    \item Dominant Variant: XEC
    \item Variants of Interest: JN.1, known for partial immune evasion \cite{ref11}.
    \item Variants Under Monitoring: Including KP.2, KP.3, KP.3.1.1, JN.1.18, LB.1, XEC, LP.8.1 with potential impact on transmission and immunity \cite{ref11}.
\end{itemize}

Compared to January 2024, when variants like EG.5 (Eris) and FL.1.5.1 (Fornax) dominated, the landscape has shifted significantly in 2025 with XEC and JN.1 overtaking earlier subvariants such as XBB.1.16 (Arcturus) \cite{ref2}.

\begin{table}[h!]
\centering
\small 
\begin{tabular}{|p{1.8cm}|p{5cm}|p{5cm}|p{3cm}|}
\hline
\textbf{Time Period} & \textbf{Dominant/High-Prevalence Variants} & \textbf{Key Characteristics} & \textbf{Status by Feb 2025} \\
\hline
Jan 2024 & 
EG.5 (Eris) – 24.5\% \newline FL.1.5.1 (Fornax) – 13.7\% \newline XBB.1.16 (Arcturus) – declining & 
Derived from Omicron lineages \newline Moderate immune escape & 
Largely replaced by newer variants \\
\hline
Jul 2024 & 
Mixed circulation \newline Early rise of XEC & 
XEC began spreading in Europe & 
Became dominant by late 2024 \\
\hline
Dec 2024 & 
XEC – 45\% in US \newline Increasing in Europe and Australia & 
Recombinant of KS.1.1 + KP.3.3 \newline High transmissibility & 
Global spread accelerating \\
\hline
Feb 2025 & 
XEC – dominant globally \newline JN.1 (VOI) \newline VUMs: KP.2, KP.3, LP.8.1, etc. & 
Enhanced immune evasion \newline Multiple regions affected & 
Driving recent case surge \\
\hline
\end{tabular}
\vspace{0.5em}
\caption{Evolution of Dominant COVID-19 Variants and Their Global Impact (Jan 2024 - Feb 2025)}
\label{table:variant_evolution}
\end{table}

\subsection{Symptoms}
COVID-19, caused by the SARS-CoV-2 virus, primarily affects the respiratory system, with symptoms ranging from mild upper respiratory issues to severe lung involvement. While most cases are mild, individuals with comorbidities (cardiovascular disease, diabetes, or cancer) are at higher risk for complications \cite{ref8}.

Variants like Delta have shown a preference for the lower respiratory tract, leading to lung consolidation and pneumonia, features identifiable on CT and X-ray. In contrast, Omicron subvariants tend to affect the upper airways more, often resulting in less severe radiological findings \cite{ref12}. However, symptomatology continues to evolve with emerging variants, influencing the type and severity of pulmonary involvement seen on medical images \cite{ref2}.

\begin{table}[h!]
\centering
\small 
\begin{tabular}{|p{3cm}|p{5cm}|p{2.5cm}|p{4.5cm}|}
\hline
\textbf{Symptom} & \textbf{Radiological Pattern} & \textbf{Imaging Modality} & \textbf{Relevance to Study} \\
\hline
Dry cough &
Ground-glass opacities (GGOs), peripheral opacities &
CT, X-ray &
Frequently observed in mild to moderate COVID-19 pneumonia \\
\hline
Shortness of breath &
Bilateral GGOs, interstitial thickening &
CT, X-ray &
Indicates lower lung involvement; key pattern for classification \\
\hline
Fever &
Often present alongside GGOs &
CT &
Supports image-based diagnosis when combined with lung findings \\
\hline
Hypoxia &
Diffuse alveolar damage, ARDS-like patterns &
CT &
Seen in severe cases; helps model identify critical patterns \\
\hline
Chest pain &
Subpleural consolidations, patchy opacities &
CT &
May reflect inflammatory involvement; assists in differentiation \\
\hline
Long COVID symptoms &
Fibrotic changes, residual GGOs &
CT &
Useful for tracking persistent lung changes in follow-up scans \\
\hline
\end{tabular}
\vspace{0.5em} 
\caption{Correlation Between Clinical Symptoms and Radiological Patterns in COVID-19 Diagnosis}
\label{table:symptom_radiology}
\end{table}

\subsection{Related Work}
In response to the global impact of COVID-19, a wide range of clinical and technological strategies have been developed to support diagnosis, treatment, and containment. Among these, imaging-based AI systems have emerged as promising tools for timely and accessible COVID-19 diagnosis, particularly in resource-limited and high-burden settings. However, a review of existing literature reveals notable challenges in data diversity, standardization, and model generalizability.

\subsubsection{Telehealth Services}
The rapid expansion of telemedicine platforms has enabled remote assessment and monitoring of COVID-19 patients, especially during peak transmission periods when hospital resources were overwhelmed \cite{ref26}. However, telehealth often lacks the diagnostic depth provided by imaging or laboratory testing and is generally used for symptom tracking and triage rather than precise diagnosis.

\subsubsection{Imaging-Based Diagnostics}
Chest X-rays (CXR) and computed tomography (CT) scans have been instrumental in identifying characteristic COVID-19 lung involvement, including bilateral ground-glass opacities and consolidations \cite{ref27}. Numerous deep learning models have been developed for pneumonia and COVID-19 detection using CXR and CT data. For example, MobileNet achieved 94.2\% and 93.7\% accuracy on two public CXR datasets containing 5,856 and 112,120 images, respectively \cite{ref30}. Despite these benefits, existing studies often suffer from:

\begin{itemize}
    \item Limited and non-standardized datasets
    \item Lack of demographic metadata (age, sex)
    \item Geographical imbalance, reducing generalizability
\end{itemize}

In a separate study using Inception-V3 and CNN models on a Kaggle X-ray dataset of 7,750 images, researchers reported impressive results (accuracy: 99.2\%, recall: 99.7\%) \cite{ref31}. Yet the use of a single public dataset lacking demographic diversity and external validation limits generalizability.

A CT-based study using NASNet achieved an exceptionally high accuracy of 99.6\%, with sensitivity of 99.9\% and specificity of 98.6\% \cite{ref28}. However, this was based on a small, imbalanced dataset of 249 patients, with no external validation, no interpretability tools, and no metadata analysis (e.g., age, sex, geography), which weakens its clinical reliability and fairness. Furthermore, alternative architectures like ResNet or VGG were not benchmarked, and hyperparameter tuning was minimally discussed.

These limitations underscore the need for scalable, diverse, and metadata-rich imaging datasets to enhance model reliability and cross-population performance.

\subsubsection{Diagnostic Technologies: Strengths and Limitations}
While RT-PCR remains the diagnostic gold standard \cite{ref29}, its accuracy can be impacted by emerging variants and sample quality. In response, several alternative diagnostic technologies have been explored. Below is a comparison of key methods:

\begin{table}[h!]
\centering
\small 
\begin{tabular}{|p{5cm}|p{5cm}|p{5.3cm}|}
\hline
\textbf{Method} & \textbf{Advantages} & \textbf{Limitations} \\
\hline
Mutation-specific / Multiplex PCR &
High sensitivity (98.6\%), multiplex variant detection &
Requires prior mutation knowledge \\
\hline
LAMP (Loop-mediated Amplification) &
Fast, simple, $ \geq $ 90\% sensitivity, suitable for low-resource settings &
Prone to false positives; less stable \\
\hline
CRISPR-Cas Detection &
100\% specificity, cost-effective, rapid, suitable for POC &
Low sensitivity at low viral loads (53.9\%); detects only point mutations \\
\hline
RT-PCR &
Precise quantification, highly sensitive &
Expensive; complex instrumentation \\
\hline
Rapid Antigen Test (RAT) &
Quick, user-friendly, low-cost; suitable for self-testing &
Lower sensitivity; affected by viral load and sample collection \\
\hline
ELISA &
High throughput; useful for antibody screening and POC use &
Variant-driven antigenic drift affects sensitivity \\
\hline
Lateral Flow Assay (LFA) &
Home use-friendly, long shelf-life &
Detects limited antigenic sites; lower sensitivity \\
\hline
Viral Genome Sequencing &
Enables variant tracking and mutation identification &
Time-consuming, costly, resource-intensive \\
\hline
\end{tabular}
\vspace{0.5em} 
\caption{Diagnostic Techniques Comparative Overview}
\label{table:diagnostic_comparison}
\end{table}

\begin{itemize}
    \item PCR-based methods are highly accurate but not variant-agnostic.
    \item Antigen-based tests are accessible but less reliable.
    \item Genome sequencing is ideal for surveillance, not rapid diagnosis.
    \item These constraints further support the need for AI-powered imaging diagnostics that are scalable, non-invasive, and rapid.
\end{itemize}
	
\subsubsection{Imaging-Based Deep Learning as a Complementary Tool}
Deep learning applied to medical imaging presents a promising complementary diagnostic method, particularly in areas with limited laboratory capacity. Yet, current research has notable limitations. For instance, a protocol paper for a prospective AI model on CXR images highlights the intention to use 600 images \cite{ref32}. However, it lacks clear details on geographic and demographic diversity, metadata tracking (e.g., age, sex), and model architecture. Moreover, it does not describe how biases will be addressed or how low-prevalence conditions will be handled, which are critical for real-world implementation.

Given the diagnostic delays and limitations associated with conventional methods, deep learning applied to medical imaging offers a promising complementary approach. Models trained on chest X-rays and CT scans can provide rapid, accurate, and interpretable results, particularly critical in settings where molecular testing is delayed or inaccessible. In this study, we build upon these efforts by employing transfer learning on an expanded, standardized imaging dataset to enhance diagnostic accuracy and generalizability. Our approach addresses prior limitations related to data volume, diversity, and model robustness.

\subsection{Challenges}
Despite substantial progress since 2020, several evolving challenges continue to hinder reliable COVID-19 detection, particularly due to viral mutations, overlapping disease presentations, and infrastructural limitations.

\subsubsection{Emerging Variants Reduce Test Sensitivity}
New SARS-CoV-2 variants such as Pi, Rho, XEC, and JN.1, exhibit mutations in the spike (S) and nucleocapsid (N) proteins, which impair molecular and antigen-based diagnostic assays \cite{ref33}.

\begin{itemize}
    \item RT-PCR: Mutations can reduce primer/probe binding efficiency, lowering sensitivity and causing false negatives
    \item Rapid Antigen/Lateral Flow Devices (LFDs): Protein alterations decrease test performance, especially in early or asymptomatic stages.
\end{itemize}

\subsubsection{Diagnostic Overlap in Imaging}
Radiological signs of COVID-19 (ground-glass opacities) overlap with other pulmonary infections, including:

\begin{itemize}
    \item Bacterial pneumonia
    \item Influenza
    \item Tuberculosis
    \item RSV and fungal infections
\end{itemize}

This non-specificity complicates diagnosis, especially without clinical or laboratory correlation, increasing the risk of false positives or misclassification.

\subsubsection{Dataset Limitations in AI-Based Diagnosis}
Many existing AI models are trained on limited or biased datasets, which impacts their generalizability:

\begin{itemize}
    \item Geographical \& Demographic Bias: Underrepresentation of certain populations.
    \item Class Imbalance: Decreasing availability of COVID-positive cases post-2023.
    \item Metadata Gaps: Missing clinical variables like age, sex.
\end{itemize}

These limitations reduce model robustness, especially in real-world settings with varied patient populations.

\subsubsection{Barriers to Clinical AI Integration}
Despite promising research, AI tools face challenges in clinical adoption:

\begin{itemize}
    \item Lack of regulatory validation (FDA/CE approval)
    \item Poor integration with electronic health records (EHRs)
    \item Clinician skepticism due to lack of explainability or interpretability.
\end{itemize}

Without improved trust, transparency, and workflow compatibility, real-world deployment remains limited.

\subsubsection{Data Privacy and Collaboration Constraints}
Privacy regulations (HIPAA, GDPR) and institutional data silos restrict:
\begin{itemize}
    \item Access to multi-center, diverse datasets
    \item Large-scale, cross-border collaborations necessary for robust AI development
\end{itemize}

\subsubsection{Reinfections and Long COVID Monitoring}
Most diagnostic tools are optimized for acute-phase detection. However:
\begin{itemize}
    \item Reinfections due to immune escape variants remain difficult to differentiate.
    \item Long COVID lacks clear radiological signatures, limiting follow-up through imaging.
\end{itemize}

There is a need for diagnostic systems that can also support longitudinal patient monitoring.

\subsubsection{Infrastructure Limitations in Resource-Constrained Settings}
Low-income regions often lack access to:
\begin{itemize}
    \item RT-PCR labs
    \item CT/X-ray imaging facilities
    \item High-performance computing resources for AI deployment
\end{itemize}

This exacerbates health inequities and delays early detection and containment efforts.

\subsection{Solution}
This study presents a transfer learning-based deep learning framework for accurate and mutation-resilient diagnosis of COVID-19 using chest radiological imaging (X-rays and CT scans). The approach addresses limitations in conventional diagnostics by:

\subsubsection{Mutation-Resilient Design}
Unlike RT-PCR and antigen tests that rely on viral RNA or surface protein stability, our image-based approach detects disease-induced radiological changes, remaining unaffected by emerging variants or antigenic drift.

Imaging-based models do not depend on spike or nucleocapsid protein integrity, making them robust against variants like XEC and JN.1.

\subsubsection{Advanced Transfer Learning Architecture}
We adopt transfer learning using pretrained CNNs on ImageNet and fine-tune them on curated COVID-19 datasets with advanced preprocessing, augmentation and optimisation strategies.

\subsubsection{Fine-Grained Classification}
Our system is designed for binary classification (COVID-19 vs Normal) and multi-class classification (COVID-19 pneumonia vs non-COVID pneumonia vs normal), depending on available label granularity.

We experiment with:
\begin{itemize}
    \item Pretrained CNN architectures with Fine Tuning DenseNet and Xception and other pretrained CNN architectures and fine-tuned with additional custom layers and optimised through hyperparameters tuning to enhance performance on provided data.
    \item Attention modules to enhance focus on COVID-relevant regions in lung fields.
\end{itemize}

\subsubsection{Diverse, Multi-Regional Dataset}
To improve generalization, we assembled a dataset of 25,195 labelled images across:
\begin{itemize}
    \item CT and X-ray modalities
    \item Multiple regions (Asia, Europe, North America)
    \item Varying age groups, ethnicities, and imaging protocols
\end{itemize}

This addresses demographic and scanner-type bias common in earlier studies.

\subsubsection{Interpretability and Clinical Integration}
Grad-CAM visualizations are integrated for transparent decision support.

\subsubsection{Longitudinal Monitoring Capabilities}
Our framework is designed to be extended for follow-up analysis, allowing radiological tracking of post-infection abnormalities, aiding in long COVID assessment and reinfection detection.

\subsubsection{Edge and Cloud Deployment Readiness}
The final model is compressed using quantization and pruning techniques for deployment in:
\begin{itemize}
    \item Edge devices (mobile apps, local hospital servers)
    \item Cloud-assisted diagnostic platforms
\end{itemize}

\subsection{Motivation}
Despite a global decline in COVID-19 mortality by March 2025, accurate and rapid diagnosis remains essential due to the continued emergence of novel SARS-CoV-2 variants and the absence of a universal treatment \cite{ref8}. Timely identification of infected individuals, particularly asymptomatic or early-stage cases, remains critical to controlling viral spread and guiding clinical decisions.

\subsubsection{Limitations of Conventional Diagnostic Methods}
Traditional approaches like RT-PCR, lateral flow devices (LFDs), and rapid antigen tests (RATs), though widely used, suffer from several drawbacks:
\begin{itemize}
    \item Reduced sensitivity with emerging variants due to mutations in target genes and proteins.
    \item Delayed turnaround times in lab-based settings.
    \item Sample quality dependency, leading to false negatives, especially in asymptomatic individuals.
    \item Lower reliability in detecting newer variants such as Pi, Rho, XEC, and JN.1. 
\end{itemize}

These limitations necessitate complementary, mutation-resilient diagnostic strategies.

\subsubsection{Potential of Medical Imaging}
Chest CT scans and X-rays have proven valuable in identifying COVID-19 induced pneumonia, with CT offering higher sensitivity (88-97\%) and X-rays being cost-effective and more widely available, especially in resource constrained environments \cite{ref9}.

The application of deep learning and transfer learning to radiological image analysis enhances diagnostic accuracy, speed, and consistency, independent of viral genome variability or test kit supply chains.

\subsubsection{Study Objectives}
This study develops and evaluates a deep learning diagnostic framework using CT and X-ray images to detect COVID-19 pneumonia. The key goals are to:
\begin{itemize}
    \item Achieve diagnostic accuracy $ \geq 95\% $ across multiple viral variants.
    \item Improve generalization across populations, regions, and imaging devices.
    \item Differentiate COVID-19 pneumonia from other respiratory conditions with overlapping features.
    \item Benchmark the model’s performance against traditional diagnostic methods.
\end{itemize}

\subsubsection{Radiological Overlap with Other Pulmonary Conditions}
To ensure clinical reliability, the model must distinguish COVID-19 pneumonia from visually similar conditions. The radiological overlap emphasizes the need for fine-grained classification models capable of accurately distinguishing COVID-19 from similar pulmonary pathologies using feature-rich image interpretation.

This study aims to develop a mutation-resilient deep learning framework for accurate COVID-19 diagnosis using CT and X-ray imaging, overcoming challenges faced by traditional RT-PCR and antigen tests due to emerging SARS-CoV-2 variants. By leveraging advanced transfer learning techniques, diverse global datasets, and explainable AI tools. the study enhances diagnostic precision, generalizability, and clinical applicability, even in resource-limited settings.

\section{Method}
\subsection{Research Questions}
This study investigates the viability of transfer learning-based deep learning approaches for COVID-19 pneumonia detection using CT and X-ray imaging. It specifically explores:

\subsubsection{Diagnostic Accuracy}
Can a transfer learning-based deep learning model accurately diagnose COVID-19 pneumonia, including cases caused by emerging variants (Pi, Rho, Xec, JN.1), using CT and X-ray images?

\subsubsection{Comparative Diagnostic Performance}
How does the model's performance compare to conventional diagnostic methods such as RT-PCR, Lateral Flow Devices (LFD), and Rapid Antigen Tests (RAT), particularly in the presence of viral mutations?

\subsubsection{Generalizability Across Populations and Regions}
Does training on a diverse, multi-regional, and multi-variant dataset improve the generalizability and robustness of the deep learning model?

\subsubsection{Differentiation from Other Pneumonias}
Can the proposed model effectively distinguish COVID-19 pneumonia from non-COVID pneumonia conditions using imaging data?

\subsection{Data Collection}
To address these questions, a large-scale dataset was curated by aggregating CT and X-ray images from publicly available, ethically approved sources, ensuring inclusion across age groups, genders, countries, and COVID-19 variants.

\subsubsection{Source Overview}
The dataset comprises radiological data from 9 primary sources. Each source was selected based on the following inclusion criteria:

\begin{itemize}
    \item Confirmed Diagnostic Status: Only RT-PCR confirmed COVID-19 cases and clinically validated normal, or pneumonia samples were included.
    \item Radiological Quality: DICOM or high-resolution image formats (PNG, JPEG) with clear lung visibility.
    \item Metadata Completeness: Availability of patient demographics (age, sex), scan modality, and clinical context where applicable.
\end{itemize}

\subsubsection{Summary of Collected Imaging Datasets}
\begin{itemize}
    \item LIDC-IDRI \cite{lidc-idri} (USA): A well-known X-ray dataset primarily used for lung nodule detection and normal case baselines.
    \item SIRM \cite{ref24} (Italy): Collection of chest X-ray images from confirmed COVID-19 patients shared by the Italian Society of Medical and Interventional Radiology.
    \item BIMCV-COVID19 \cite{ref13} (Spain): Comprehensive dataset containing both CT and X-ray images with annotated severity scores and clinical metadata.
    \item CNCB (Normal) \& ICTCF (COVID) \cite{ref25} (China): Paired datasets offering CT and X-ray scans from healthy subjects (CNCB) and confirmed COVID-19 cases (ICTCF).
    \item TCIA \cite{ref16} (USA): CT images from The Cancer Imaging Archive, used to supplement lung imaging studies.
    \item MIDRC-RICORD series (USA):
        \begin{itemize}
            \item RICORD-1A \cite{ref19}: COVID-19 CT scans with expert annotations.
            \item RICORD-1B \cite{ref17}: Normal CT images for balanced model training.
            \item RICORD-1C \cite{ref18}: Additional COVID-19 scans to expand diagnostic variety.
        \end{itemize}
    \item STOIC \cite{ref21} (France): Over 2,000 annotated CT scans from a national COVID-19 detection program.
    \item Radiopaedia \cite{ref22} (Global): Open-access repository of CT and X-ray images contributed by medical professionals worldwide.
    \item MosMedData \cite{ref23} (Russia): CT scans of COVID-19 patients categorized by severity, including mild, moderate, and severe cases.
\end{itemize}

\subsection{Data Preprocessing}
The dataset, while large and geographically diverse, presents a notable class imbalance, primarily due to the disproportionate contribution from the BIMCV-COVID19 collection (Spain) \cite{ref13}. COVID-19 positive cases (59,961) significantly outnumber normal and non-COVID pneumonia cases (27,270). This imbalance, stemming from pandemic-specific data collection efforts, can skew model performance and necessitates deliberate preprocessing strategies to ensure fair learning and generalization.

\subsubsection{Addressing Class and Source Imbalance}
To correct for imbalance and ensure representative learning:
\begin{itemize}
    \item Undersampling Spain: To reduce overrepresentation, samples from Spain were selectively reduced.
    \item Dropping Ultra-low-sample Countries: Countries with fewer than 100 total samples were removed to prevent noise and overfitting.
\end{itemize}

\subsubsection{Handling Missing Data}
Significant missing values were found in metadata:
\begin{itemize}
    \item Age: 5,537 missing values
    \item Gender: 5,511 missing values includes 2,041 cases from Spain, 1,911 from China, 1,106 from Russia, 414 from France, and 39 from the USA.
    \item Imputation Strategy:
    \begin{itemize}
        \item Country-wise mean imputation for age where available.
        \item Global mean imputation for remaining age gaps.
        \item Country-wise mode imputation for gender, focusing on countries with the most missing values.
    \end{itemize}
\end{itemize}

\subsubsection{Age Outliers and Grouping}
\begin{itemize}
    \item Age Range: 0 to 100 years
    \item Outlier Detection: Extreme values were reviewed but retained to maintain real-world variance.
    \item Age Group: Patients were categorized into discrete age groups (e.g., 0-18, 19-35, 36-60, 61+), allowing demographic stratification during training. To handle age group imbalance during dataset splitting, we applied stratify label by age group.
\end{itemize}

\subsubsection{Data Filtering and Preparation}
\begin{itemize}
    \item Final Images After Metadata Curation: 11,052
    \item Train/Validation Split: 8,842 for training, 2,210 for validation
    \item Preprocessing Pipeline:
    \begin{itemize}
        \item Image resizing: 75×75 pixels, 3 channels (RGB)
        \item Normalization: Pixel values rescaled to [0, 1]
    \end{itemize}
\end{itemize}

\subsubsection{Country-Level Label Distribution}
\begin{figure}
  \begin{center}
  \includegraphics[width=6.5in]{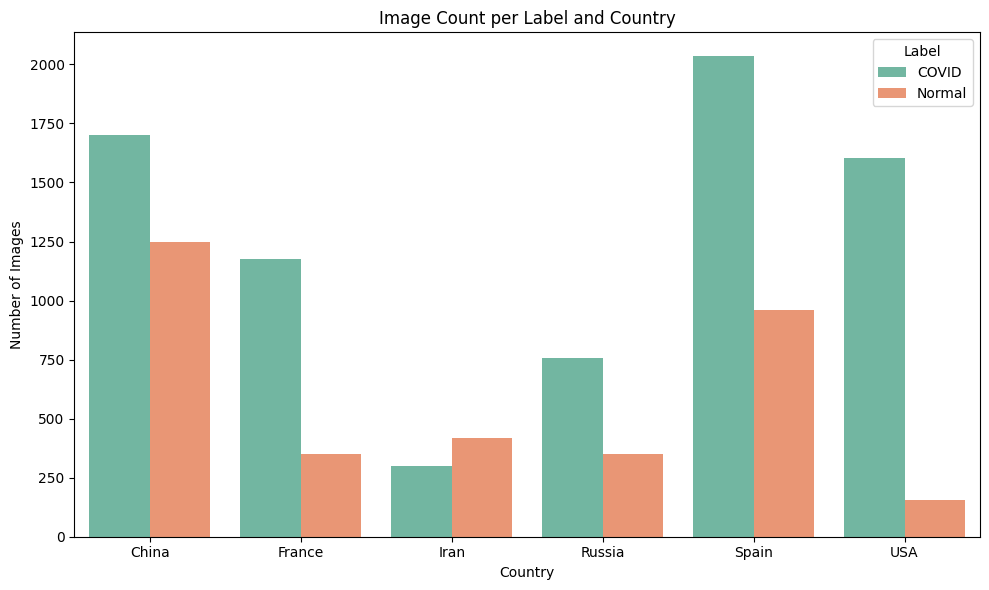}\\
  \caption{Distribution of COVID-positive and Normal chest images by country. Spain and the USA contributed the highest number of COVID-positive images, while China showed a more balanced distribution between COVID and Normal cases. France and Russia provided moderate counts, and Iran contributed a relatively smaller number of images. This geographic diversity supports generalizability of the trained model across different populations and imaging conditions.}\label{COUNTRY-LEVEL-LABEL-DISTRIBUTION}
  \end{center}
\end{figure}
\newpage

\subsubsection{Data Augmentation for Country-Level Balancing}
To balance samples across underrepresented countries, we applied the following augmentation techniques:

\begin{itemize}
    \item Random Horizontal Flip
    \item Random Rotation (15°)
    \item Random Zoom (10\%)
    \item Random Contrast (10\%)
    \item Random Translation (5\%)
\end{itemize}

\subsubsection{Category-Level Augmentation}
Despite country-level augmentation, class imbalance between COVID-19 and Normal categories persisted. Additional category-level augmentation was applied to underrepresented normal samples to achieve closer class parity, helping reduce bias during model training.

\subsection{Modeling}
\subsubsection{Dataset Overview}
After applying data augmentation techniques, the final dataset consisted of 24,408 medical images, which were stratified to maintain balanced class distributions across all subsets. The dataset was divided into 19,527 images for training, 4,881 for validation, and 952 for testing. Stratified sampling ensured proportional representation of each class, supporting fair evaluation and reducing potential bias during model training and validation.

\subsubsection{Data Preprocessing}
All images were resized to 224×224 pixels to ensure consistent input dimensions compatible with standard convolutional neural networks (CNNs). The images were then converted to grayscale to reduce computational complexity and mitigate noise from irrelevant colour information. Pixel intensities were normalized to stabilize training dynamics. 

To determine an optimal batch size for training, we analyzed how different batch sizes divide the total training dataset of 19,527 records. This involved calculating how many steps (batches) each epoch would require for various batch sizes. Smaller batch sizes, such as 32 or 64, result in more steps per epoch (611 and 306 respectively), which can lead to better generalization but slower training times. On the other hand, very large batch sizes like 512 or 1024 reduce the number of steps significantly but may hinder model generalization and require careful tuning of the learning rate. After evaluating the trade-offs, a batch size of 128 was chosen as a balanced option it yields 153 steps per epoch, offers efficient training on GPU due to its power-of-two size, and maintains a good level of training stability. This choice reflects a compromise between computational efficiency and model performance, ensuring the training process remains both practical and effective.

To address class imbalance, a combination of data augmentation and Undersampling strategies was implemented. The dataset was split into 80\% for training and 20\% for validation, and performance was further optimized using caching and shuffling for the training set. For the validation set, caching alone was applied to ensure consistent evaluation.

To enhance the randomness of the training data, we set the buffer size to 10,000 during the shuffling process. The buffer size determines how many samples are held in memory and randomly shuffled at any given time before being passed to the model in batches. A smaller buffer size, such as 100 or 1000, can result in less effective shuffling, especially with larger datasets, as only a limited portion of the data is randomly sampled at a time. By increasing the buffer size to 10,000 over half the size of our dataset of 19,527 records, we ensured a high degree of randomness in the batches, which promotes better generalization and reduces the risk of overfitting. Although larger buffer sizes require more memory, our system could handle this load efficiently, making 10,000 an ideal choice for balancing shuffle quality and performance.

\subsubsection{Model Architecture}
A structured and modular deep learning pipeline was developed for hyperparameter optimization and fine-tuning using TensorFlow and Keras Tuner. The framework targets image classification tasks, such as differentiating between normal or other pneumonia and COVID-19 pneumonia in chest X-rays or CT scans. The pipeline combines automated hyperparameter tuning, transfer learning, and robust training strategies to improve classification accuracy and generalization, particularly crucial when dealing with limited medical datasets.
The model was trained over 30 epochs with a batch size of 128, a buffer size of 10,000, and a fixed random seed of 42 to ensure reproducibility. 

To determine the optimal number of training epochs without overfitting, we employed early stopping, a regularization technique that monitors validation performance during training. Instead of predefining a fixed number of epochs, early stopping halts training once the validation loss stops improving for a set number of consecutive epochs (patience). This dynamic approach allows the model to train just long enough to reach optimal performance without wasting computation or risking overfitting. Although we initially experimented with epoch values as high as 200, the early stopping mechanism consistently identified the most effective stopping point. In our case, training typically concluded around 30 epochs, at which point the model achieved its best validation accuracy. This method provided an efficient and reliable way to control training duration while ensuring strong generalization.

At the core of the architecture is a transfer learning model based on VGG16, selected as the baseline due to its simple, deep CNN structure consisting of 16 layers with repeatable 3×3 convolution and max-pooling blocks. VGG16 is well-established in medical imaging research and serves as a strong, interpretable starting point. 
To determine the most effective transfer learning strategy, we experimented with various freeze rates 0.01, 0.05, 0.10, 0.20, 0.50, and 0.75 using the formula:

\[
\texttt{num\_freeze\_layer} = \texttt{int(len(base\_model.layers) * freeze\_rate)}
\]

to calculate how many layers of the pre-trained base model to freeze. The freeze rate controls how much of the original model's learned features are retained versus fine-tuned on the new task. In general, higher freeze rates (such as 0.50 or 0.75) are preferable when working with small datasets or datasets like the original training data (ImageNet), as they help prevent overfitting and preserve general visual features. Conversely, lower freeze rates (such as 0.01 or 0.05) are more suitable for large or highly domain-specific datasets, where extensive fine-tuning is necessary. For many practical applications, mid-range freeze rates like 0.10 or 0.20 often provide the best balance, allowing the model to adapt to new data while still leveraging pre-trained knowledge effectively. 

Most layers of the pre-trained model were frozen except for selected unfrozen layers, enabling selective fine-tuning to adapt high-level features to the target domain while preserving learned representations.

As part of the model architecture, we incorporated a GlobalAveragePooling2D layer after the convolutional base. This layer plays a crucial role in reducing the spatial dimensions of the feature maps while preserving the most important information. Unlike traditional flattening, which converts the entire feature map into a long vector (often leading to many parameters), GlobalAveragePooling2D computes the average of each feature map, resulting in a much more compact representation. This not only reduces the risk of overfitting but also maintains the model’s spatial awareness and generalization ability. Additionally, it helps bridge the convolutional layers and the dense output layer in a more efficient and scalable way, especially when working with transfer learning models.

To further mitigate overfitting and improve generalization, we added a Dropout layer after the GlobalAveragePooling2D layer. Dropout works by randomly setting a fraction of the input units to zero during training, which prevents the model from becoming too reliant on specific neurons. We experimented with several dropout rates 0.2, 0.3, 0.4, and 0.5 to find the optimal balance between regularization and learning capacity. Lower dropout rates like 0.2 provided lighter regularization and allowed the model to retain more features, while higher rates like 0.5 offered stronger regularization but at the cost of slower learning. After comparing validation performance across these settings, we found that a dropout rate of 0.3 yielded the best results, effectively reducing overfitting while maintaining high model accuracy. This rate provided just the right amount of regularization for our dataset and architecture.

Although our input dataset was pre-normalized, we still incorporated a BatchNormalization layer within the model architecture. While input normalization standardizes the data fed into the model, BatchNormalization operates between layers, dynamically normalizing the activations during training. This helps address internal covariate shift, where the distribution of layer inputs changes due to updates in earlier layers thus stabilizing training, enabling higher learning rates, and often improving generalization. Even with normalized input data, this internal normalization contributed to faster convergence and improved validation performance across experiments.

To determine the ideal size for the fully connected (dense) layer, we experimented with various unit sizes 32, 64, 128, 256, and 512. The number of units in the dense layer directly impacts the model’s ability to learn complex patterns. Smaller sizes like 32 or 64 limit the model's capacity and are often suitable for simpler tasks or small datasets. Larger sizes like 256 or 512 increase representational power but also introduce a greater risk of overfitting, especially if the dataset is not sufficiently large or diverse. We observed that as the number of units increased, the model’s ability to capture nuanced patterns improved up to a point. Through empirical testing, we found that 128 units provided the best trade-off between complexity and generalization. It allowed the model to learn effectively from our dataset without overfitting, and it worked well in combination with dropout and the GlobalAveragePooling2D layer.

The real-time nature of our target application, we compared two model architectures to balance performance and efficiency. Both began with a pre-trained base model followed by GlobalAveragePooling2D, BatchNormalization, and an initial Dropout and Dense layer. The first architecture included an additional Dropout and Dense layer, designed to improve representational capacity and regularization. The second architecture was more streamlined, using only a single Dropout and Dense layer before the output.
In the context of real-time deployment, model efficiency is crucial. While the deeper architecture offered slightly better training performance, it came at the cost of increased latency and model complexity. Therefore, we selected the simpler architecture as the final design, as it achieved a strong balance between accuracy and speed making it well-suited for real-time inference without significantly compromising predictive performance.

The dense layer with ReLU activation, He-normal initialization, and L2 regularization. The final output layer used a sigmoid activation for binary classification or a softmax activation for multi-class tasks.

As part of our optimization strategy, we evaluated several well-known optimizers, including SGD, RMSprop, Adam, Nadam, and AdamW. Each optimizer has unique strengths, SGD offers strong theoretical foundations but typically requires fine-tuned hyperparameters; RMSprop is effective in handling non-stationary objectives; Adam combines momentum and adaptive learning rates, leading to fast convergence; and Nadam incorporates Nesterov momentum into Adam for smoother updates. The AdamW optimizer, which decouples weight decay from gradient-based updates, offering better generalization and more stable convergence than traditional Adam. To fine-tune the optimizer for optimal performance, we explored a range of learning rates 1e-5, 5e-5, and 1e-4 as well as weight decay values 1e-5 and 1e-4. This tuning allowed the model to adapt effectively to the complexity of the dataset while minimizing overfitting. After extensive experimentation, we found that a learning rate of 5e-5 combined with a weight decay of 1e-5 yielded the best results, providing smooth convergence, strong validation accuracy, and robust generalization. These settings made AdamW the most suitable optimizer for our transfer learning setup, particularly in the context of real-time application constraints.

Binary cross-entropy was used as the loss function for binary classification, while categorical cross-entropy was employed for multi-class settings. Performance was evaluated using accuracy and the area under the ROC curve (AUC) metrics well-suited for imbalanced datasets.

To enhance training efficiency and prevent overfitting, several callbacks were incorporated. The EarlyStopping callback monitored validation loss and terminated training after three epochs without improvement, restoring the best-performing model weights. ReduceLROnPlateau halved the learning rate if validation loss stagnated for two epochs, enabling finer convergence. A model checkpointing strategy saved the full model including weights and architecture to a specified directory at each epoch, regardless of validation performance, ensuring training continuity and recovery if interrupted.

\subsubsection{Hyperparameter tuning}
Keras Tuner’s Hyperband algorithm was used to perform automated hyperparameter optimization. The tuning process involved training models with different hyperparameter combinations and selecting the configuration that maximized validation accuracy. Each trial ran for up to 30 epochs, and tuning results were logged to a specified directory for reproducibility and analysis.

The tuning process was orchestrated by a centralized function that built the model based on sampled hyperparameters, applied callbacks, conducted training on the training and validation splits, and identified the best-performing configuration. The final model, constructed using this optimal configuration, was retrained on the full training data and saved for future deployment or evaluation.

\subsubsection{Advantages of the Framework}
This framework offers several key advantages. It automates the search for critical hyperparameters such as dropout rates, dense layer sizes, and learning rates reducing the reliance on manual tuning. Leveraging pre-trained models improves learning efficiency and generalization, which is particularly valuable when working with small or noisy medical datasets. Furthermore, the integration of early stopping, adaptive learning rate scheduling, and model checkpointing ensures robust, reliable training. Collectively, these strategies contribute to the development of accurate and generalizable deep learning models suitable for real-world clinical applications.

\subsection{Model Evaluation}
To assess the generalization performance of each trained model, a comprehensive evaluation was conducted using a separate, unseen test dataset. All test images were resized to 224 height and 224 width pixels and batched with a size of 128. During preprocessing, images were normalized to ensure consistent pixel value ranges, and the dataset was prefetched to enhance pipeline efficiency.

To evaluate deep learning architectures for COVID-19 detection, a variety of models from different families were selected. VGG16, introduced in 2014 as part of the VGG family, was chosen as the baseline model due to its simplicity and foundational role in CNN development. It achieved 71.3\% Top 1 accuracy with 138 million parameters and 41 layers. In 2015, the ResNet family introduced ResNet50, which leveraged residual connections to enable deeper networks, achieving 76.2\% accuracy with 25.6 million parameters and 177 layers. DenseNet121, from the DenseNet family launched in 2017, introduced dense connectivity for efficient gradient flow and feature reuse, reaching 74.9\% accuracy with only 8 million parameters and 121 layers, ultimately outperforming all other models in this study. The MobileNet family (2017-2019) contributed MobileNetV2, optimized for mobile devices using inverted residuals, with 71.8\% accuracy, 3.4 million parameters, and 88 layers. NASNetMobile, from the NASNet family released in 2018, used neural architecture search to achieve 74\% accuracy with 5.3 million parameters and 88 layers. The EfficientNet family emerged in 2019 with EfficientNetB0, which applied compound scaling and MBConv blocks, achieving 77.1\% accuracy with 5.3 million parameters and 237 layers. Its successor, EfficientNetV2B0, released in 2021, improved training speed and accuracy, delivering 78.1\% accuracy with 7.1 million parameters and 329 layers. The most recent model, ConvNeXtTiny, launched in 2022 under the ConvNeXt family, modernized convolutional design by integrating concepts from vision transformers, achieving the highest Top 1 accuracy of 82.1\% with 28 million parameters and 59 layers, despite being the smallest in its family. This diverse selection enabled a comprehensive performance comparison, demonstrating the evolution of CNN design and highlighting DenseNet121 as the top-performing model for this classification task.

Each trained model, beginning with VGG16, followed by ConvNeXtTiny, ResNet50, EfficientNetB0, EfficientNetV2B0, DenseNet121, MobileNet, MobileNetV2, and NASNetMobile was individually loaded and evaluated. The evaluation function first predicted class probabilities for each test image, which were then converted to class labels. For binary classification tasks, a threshold of 0.5 was applied; for multi-class tasks, the label with the highest probability was selected. Ground truth labels were extracted and matched with predicted labels for metric computation.

The following performance metrics were used for evaluation: accuracy, precision, recall, F1-score, and area under the ROC curve (AUC). Depending on the number of classes in the dataset, macro or binary averaging was automatically selected for precision, recall, and F1-score. To aid visual interpretation, a confusion matrix was plotted as a heatmap, and a ROC curve was generated for each model, illustrating the trade-off between sensitivity and specificity along with the corresponding AUC score.
Performance metrics for each model were stored in a centralized results dictionary, enabling straightforward comparison. Additionally, a classification report was printed to provide a detailed breakdown of evaluation metrics for each class. Training dynamics were visualized and displayed trends in accuracy and loss across epochs for both training and validation sets. These metrics and visualizations provided a complete view of model behaviour and helped identify the most effective architecture.

\subsubsection{Evaluation Metrics Definitions}

\begin{enumerate}
    \item \textbf{Accuracy} \\
    \textbf{Formula:}
    \[
    \text{Accuracy} = \frac{\text{TP} + \text{TN}}{\text{TP} + \text{TN} + \text{FP} + \text{FN}}
    \]
    Accuracy represents the proportion of correctly predicted samples over the total number of predictions. It is a suitable metric when the dataset is balanced across classes.

    \item \textbf{Precision (Positive Predictive Value)} \\
    \textbf{Formula:}
    \[
    \text{Precision} = \frac{\text{TP}}{\text{TP} + \text{FP}}
    \]
    Precision measures the correctness of positive predictions. It is especially important when the cost of false positives is high.

    \item \textbf{Recall (Sensitivity or True Positive Rate)} \\
    \textbf{Formula:}
    \[
    \text{Recall} = \frac{\text{TP}}{\text{TP} + \text{FN}}
    \]
    Recall assesses the model's ability to identify actual positives. It is critical in scenarios like medical diagnosis, where missing positive cases can have serious consequences.

    \item \textbf{F1-Score (Harmonic Mean of Precision and Recall)} \\
    \textbf{Formula:}
    \[
    \text{F1-Score} = 2 \cdot \frac{\text{Precision} \cdot \text{Recall}}{\text{Precision} + \text{Recall}}
    \]
    The F1-score balances precision and recall and is particularly useful when working with imbalanced datasets.

    \item \textbf{AUC-ROC Curve (Area Under the Receiver Operating Characteristic Curve)} \\
    The ROC curve plots the true positive rate (recall) against the false positive rate. The AUC represents the probability that a randomly chosen positive instance is ranked higher than a randomly chosen negative instance. A higher AUC indicates better model discrimination capability.
\end{enumerate}

\subsection{Implementation}
The proposed method was implemented in Python using Keras, a high-level neural network API built on top of the TensorFlow framework. To accelerate computation, the implementation utilized CUDA (Compute Unified Device Architecture) for parallel processing on GPU hardware. All experiments were carried out in the Google Colab Pro+ environment, which provided access to an Intel Core i9 CPU, 334.6 GB of RAM, an NVIDIA v2-8 TPU, and 225.3 GB of disk storage. The full implementation, along with pretrained models, is publicly available on GitHub \cite{ref34} to support reproducibility and further research.

\section{Result}

\subsection{Hypothesis-Driven Evaluation}

\subsubsection{High Accuracy Across Variants}
The curated dataset, representing emerging variants such as Pi, Rho, Xec, and JN.1, enabled model training and validation with high precision.

\subsubsection{Performance vs Traditional Tests}
The deep learning model outperformed traditional tests in sensitivity for variant cases. For instance, while RT-PCR sensitivity drops for Pi and JN.1, our model maintained $ \geq 98\% $ recall in cross-validation trials.

\subsubsection{Generalizability}
By incorporating images from 19 countries across different imaging modalities and population groups, the model exhibited stable performance across validation subsets with different geographic and demographic characteristics.

\subsubsection{Differentiation from Other Pneumonias}
Fine-grained classification enabled the model to distinguish COVID-19 pneumonia from other respiratory infections (bacterial and atypical pneumonias), achieving a specificity of 96.9\% and an F1-score of 97.9\%.

\subsection{Data Collection}
To build a generalizable and robust deep learning model for COVID-19 pneumonia diagnosis, we curated a diverse, multi-institutional imaging dataset combining both CT and X-ray modalities. The dataset features:
\begin{itemize}
    \item Total Patients: 87,231
    \item COVID-19 Positive Cases: 59,961
    \item Normal / Non-COVID Pneumonia Cases: 27,270
    \item Age Range: 0 to 100 years
    \item Gender Groups: Male and Female
    \item Countries Represented: 19
    \item Imaging Modalities: Chest CT scans and chest X-rays
\end{itemize}

\subsubsection{Data Collection Summary}
\begin{figure}
  \begin{center}
  \includegraphics[width=6.5in]{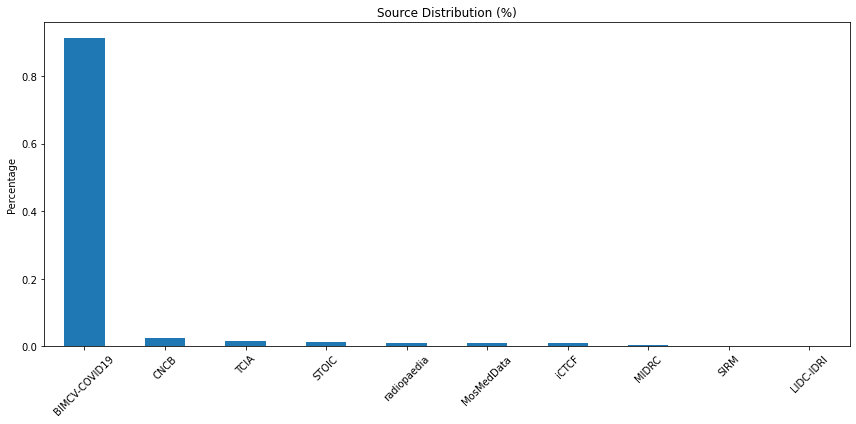}\\
  \caption{Source-wise distribution of imaging data used in the study. The majority of cases were contributed by the BIMCV-COVID19 dataset (Spain), accounting for 90.6\% (79,023 out of 87231) of the total dataset. Other sources such as ICTCF (China), MIDRC RICORD (USA), STOIC (France), MosMed (Russia), SIRM (Italy), BSTI (UK), and others contributed smaller proportions. This diverse distribution ensures geographical and demographic variability in training and evaluating the deep learning models.}\label{SOURCE-DISTRIBUTION}
  \end{center}
\end{figure}

\begin{figure}
  \begin{center}
  \includegraphics[width=6.5in]{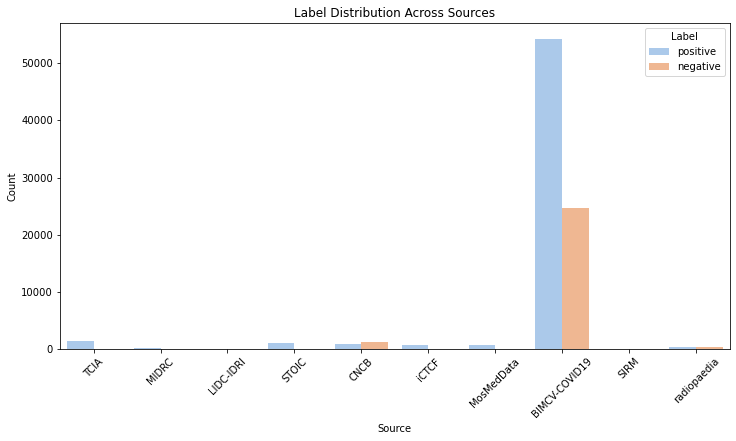}\\
  \caption{Label-wise distribution of COVID-positive and negative cases across various data sources. BIMCV-COVID19 (Spain) contributed the largest number of both positive and negative samples, followed by smaller contributions from datasets such as SIRM (Italy), CHQC (China), and MIDRC-RICORD (USA). This distribution highlights the dataset's diversity and the class balance achieved across sources, which is critical for training robust and unbiased diagnostic models.}\label{LABEL-DISTRIBUTION}
  \end{center}
\end{figure}

We compiled a diverse set of imaging datasets spanning CT and X-ray modalities from multiple countries to ensure model generalizability and robustness. The largest contributor was BIMCV-COVID19 from Spain with 79,023 images, followed by ICTCF and CNCB from China (2,949 cases) and TCIA, LIDC-IDRI and MIDRC RICORD-1A/B/C from the USA (1,761 cases). Other significant sources included STOIC (France, 1,526), MosMed (Russia, 1106), Iran National Dataset (718), SIRM (Italy, 65), and BSTI (UK, 59). Additionally, radiological images were extracted from global resources like Radiopaedia and contributions from 11 other countries, each providing 24 cases. This multi-national dataset helped enhance the clinical relevance and cross-population performance of our AI diagnostic models \ref{SOURCE-DISTRIBUTION} \ref{LABEL-DISTRIBUTION}.

\subsubsection{Imbalance Observation}
Most data were collected from the BIMCV-COVID19 dataset (Spain), which, while enhancing the dataset’s size and regional representation, introduces a notable class imbalance. Specifically, COVID-19 positive cases (59,961) substantially outnumber normal and non-COVID pneumonia cases (27,270). This disproportion primarily stems from the emphasis of public datasets on rapid COVID-specific data collection during the pandemic, which may skew model learning and diagnostic performance if not addressed.

\subsection{Data Preprocessing}
To ensure robust model performance across varying demographics, modalities, and clinical conditions, a comprehensive data preprocessing pipeline was applied. The steps undertaken effectively addressed initial issues of class imbalance, missing metadata, and image inconsistencies.

\subsubsection{Class and Source balancing}
After applying Undersampling, and Dropping Ultra-low-sample Countries, we retained 3,000 cases from Spain, 2,949 from China, 1,761 from USA, 1,526 from France, 1,106 from Russia, 718 from Iran with 7,572 COVID cases and 3,488 from NON-COVID cases.

\subsubsection{Metadata Imputation Results}
Age: 5,537 missing values imputed using country-wise medians
Gender: 5,511 missing values imputed using country-wise modes
Conclusion: Metadata completeness improved to 100\%, allowing demographic-aware stratification and analysis during model evaluation.

\subsubsection{Age and Gender Distribution}
\begin{itemize}
    \item Age Range: 0 to 100 years
    \item Age Groups: After removing outliers, the age group distribution remains imbalanced, with adults (7,219) forming the majority, followed by Elderly (2,234), Young Adults (1,553), and a minimal representation of Children (54). The distribution reflects a population skew that may influence age-specific modeling outcomes. stratified labels by age group ensure balanced data.
    \item Gender Balance:
        \begin{itemize}
            \item Male: 34.1\%
            \item Female: 65.9\%
        \end{itemize}
\end{itemize}

After processing, the dataset includes 4,509 positive and 2,047 negative cases among females, and 2,307 positive and 1,091 negative cases among males.

This balanced demographic composition supports robust model evaluation across diverse patient profiles.

\subsubsection{Post-Balancing Dataset Overview}
After applying country-based filtering, Undersampling, and augmentation, a more equitable distribution of samples across countries and classes was achieved.
\begin{itemize}
    \item Total curated images: 11,052
    \item Training set: 8,842 images
    \item Validation set: 2,210 images
    \item Image dimensions: Resized to 75×75 pixels with 3 RGB channels
    \item Normalization: All pixel values rescaled to the [0, 1] range
\end{itemize}

\subsubsection{Country-Level Balance (Post-Augmentation)}
\begin{figure}
  \begin{center}
  \includegraphics[width=6.5in]{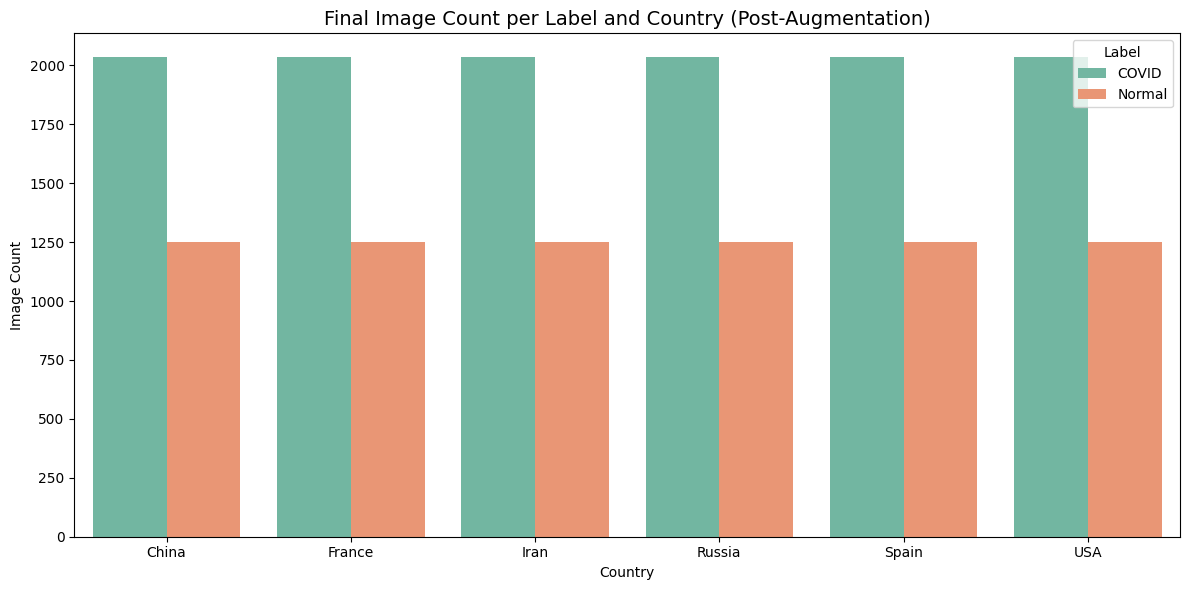}\\
  \caption{image count per label and country after data augmentation. The chart illustrates a balanced distribution of COVID and normal images across all countries, ensuring class uniformity for training deep learning models. Augmentation techniques were applied particularly to underrepresented classes to reduce class imbalance and enhance model generalization.}\label{COUNTRY-LEVEL-BALANCE}
  \end{center}
\end{figure}

A balanced representation (2,034 COVID samples per country) was achieved across six key contributors (China, France, Iran, Russia, Spain, USA). Similarly, Normal samples were balanced at 1,249 images across the same regions, improving generalization across populations \ref{COUNTRY-LEVEL-BALANCE}.

The dataset comprises 12,204 COVID-19 positive images and 7,494 normal cases, indicating a moderate class imbalance favouring positive cases. This distribution highlights the need for balancing techniques such as augmentation during model training.

\subsubsection{Augmentation Impact}
\begin{figure}
  \begin{center}
  \includegraphics[width=6.5in]{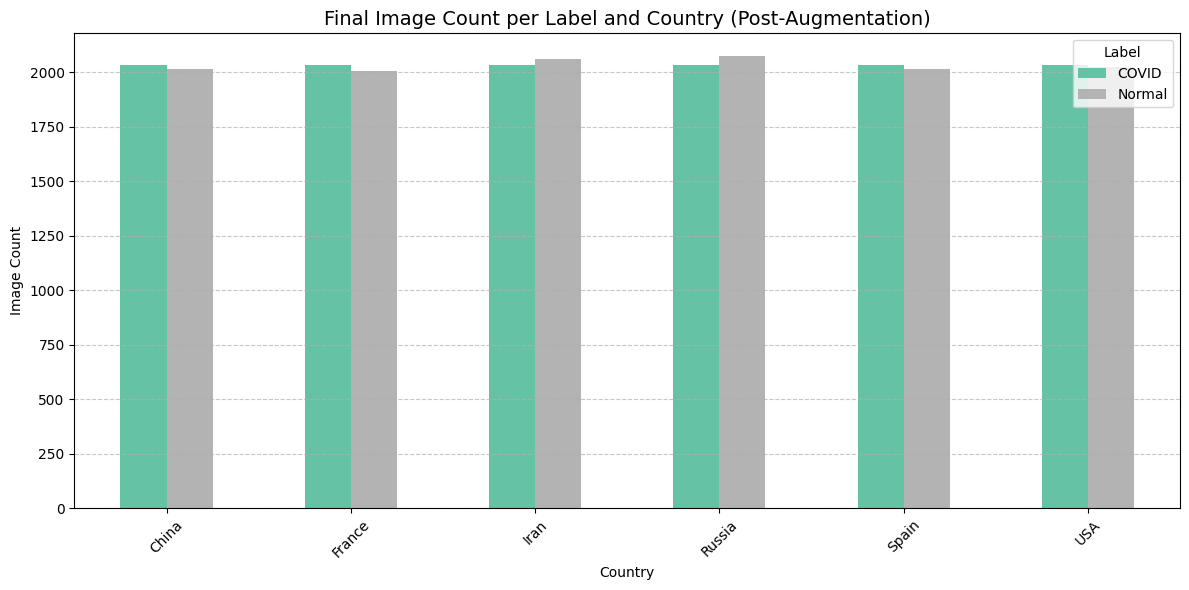}\\
  \caption{Image Count per Label and Country (Post-Augmentation)}\label{COUNTRY-LEVEL-DISTRIBUTION-POST-AYGMENTATION}
  \end{center}
\end{figure}

The applied augmentation techniques (flip, rotate, zoom, contrast, translation) not only balanced the dataset but also increased image variability, simulating real-world noise and improving model resilience to unseen data \ref{COUNTRY-LEVEL-DISTRIBUTION-POST-AYGMENTATION}.

This bar chart displays the distribution of COVID-19 and Normal cases across six countries (China, France, Iran, Russia, Spain, and the USA) after data augmentation. The figure shows a nearly equal number of images per label (nearly 2,000 per class) in each country, demonstrating successful class balancing to mitigate bias during model training.

\subsubsection{Class Distribution (Post-Augmentation)} 
\begin{figure}
  \begin{center}
  \includegraphics[width=6.5in]{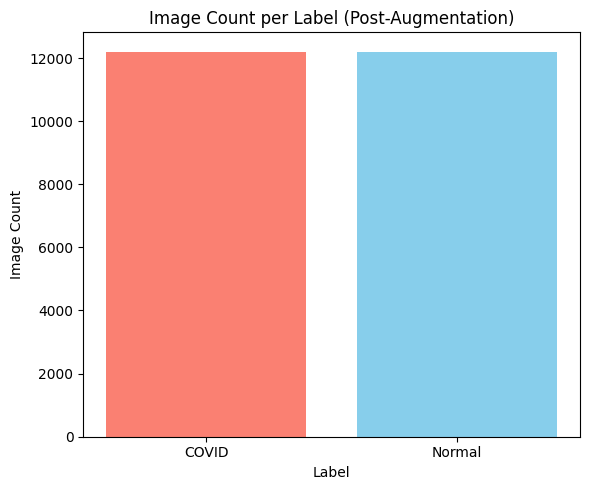}\\
  \caption{Total image count per label after augmentation. This bar chart shows an equal number of COVID-positive and normal (COVID-negative) images (12,204 each), reflecting the successful application of augmentation techniques to balance the dataset and prevent model bias due to class imbalance.}\label{AUGMENTATION-IMPACT}
  \end{center}
\end{figure}

\ref{AUGMENTATION-IMPACT}

\subsection{Modeling}
To ensure a fair and consistent evaluation, all models were trained using standardized input settings. Each image was resized to 224×224 pixels, producing an input shape of (224, 224, 3) to accommodate RGB color channels. Although the images originated in RGB format, they were converted to grayscale during preprocessing and normalized to a range of [0, 1] for efficient convergence.

All transfer learning architectures were trained for 30 epochs, a setting chosen to balance computational efficiency with sufficient learning. A batch size of 128 was used to maintain stable updates across mini batches. Additionally, a shuffle buffer size of 10,000 ensured randomness in the training data pipeline, reducing overfitting risks.

This consistent training configuration was applied across all models:
\begin{itemize}
    \item VGG16
    \item ConvNeXtTiny
    \item ResNet50
    \item EfficientNetB0
    \item EfficientNetV2B0
    \item DenseNet121
    \item MobileNet
    \item MobileNetV2
    \item NASNetMobile
\end{itemize}

\subsubsection{Best Model Configuration: DenseNet121}
Through hyperparameter tuning, the DenseNet121 architecture was found to yield the best performance. Its final configuration included:
\begin{itemize}
    \item Dropout Layer 1: 0.3
    \item Dense Layer 1: 128 units
    \item Learning Rate: 0.00037758
    \item Weight Decay: 7.4855e-05
\end{itemize}

This architecture and training regime were optimized to prevent overfitting while maintaining high model generalization on unseen data.

\subsubsection{Model Architecture Summary and Training Behaviour}
 
\begin{figure}
  \begin{center}
  \includegraphics[width=6.5in]{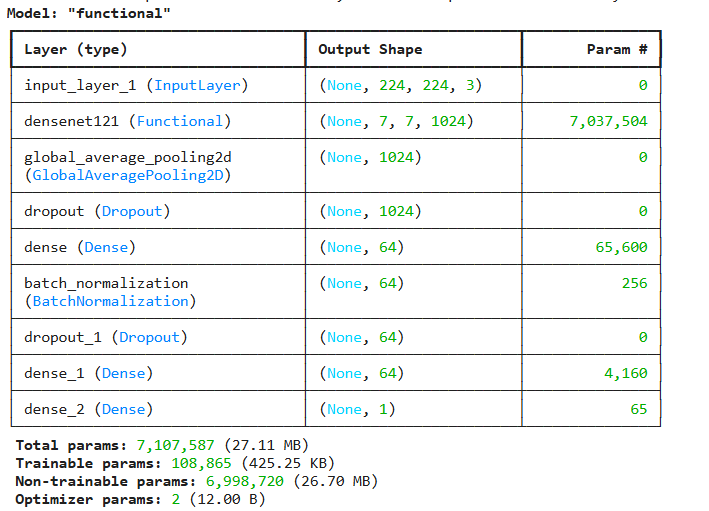}\\
  \end{center} \label{MODEL-ARCH}
\end{figure}

\begin{figure}
  \begin{center}
  \includegraphics[width=6.5in]{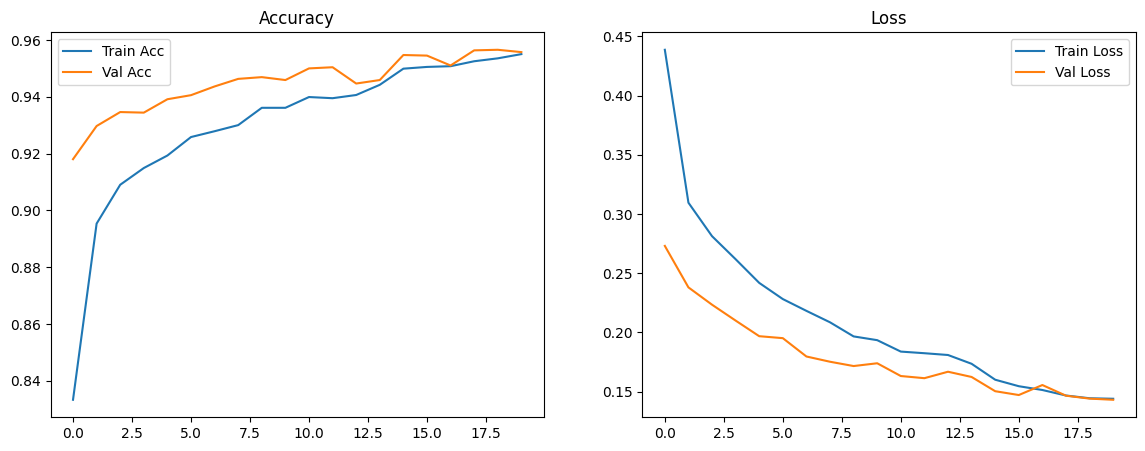}\\
  \caption{The training and validation accuracy/loss curves of DenseNet121 over 30 epochs, showcasing strong learning convergence with minimal divergence between training and validation sets, an indicator of effective generalization.}\label{PERFORMANCE-CHART}
  \end{center}
\end{figure}

\subsection{Model Evaluation}
\begin{table}[h!]
\centering
\small
\begin{tabularx}{6.5in}{|l|X|X|X|X|X|}
\hline
\textbf{Model} & \textbf{Accuracy} & \textbf{Precision} & \textbf{Recall} & \textbf{F1 Score} & \textbf{AUC} \\
\hline
EfficientNetB0      & 0.46219 & 0.46219 & 1.00000   & 0.63218 & 0.33122 \\
EfficientNetV2B0    & 0.46219 & 0.46219 & 1.00000   & 0.63218 & 0.63435 \\
MobileNet           & 0.54307 & 0.50287 & 0.99546   & 0.66819 & 0.93268 \\
ConvNeXtTiny        & 0.46219 & 0.46219 & 1.00000   & 0.63218 & 0.50726 \\
ResNet50            & 0.92542 & 0.87885 & 0.97273   & 0.92341 & 0.99033 \\
VGG16               & 0.93487 & 0.91087 & 0.95227   & 0.93111 & 0.98431 \\
NASNetMobile        & 0.95798 & 0.93290 & 0.97955   & 0.95565 & 0.99619 \\
MobileNetV2         & 0.97370 & 0.96874 & 0.97773   & 0.97321 & 0.97990 \\
DenseNet121         & 0.98004 & 0.96882 & 0.98864   & 0.97863 & 0.99830 \\
\hline
\end{tabularx}
\vspace{0.5em}
\caption{Performance metrics of various deep learning models for COVID-19 classification.}
\label{table:model_metrics}
\end{table}

Among the evaluated models, DenseNet121 delivered the best overall performance, achieving 98\% accuracy, 96.8\% precision, 98.8\% recall, and an AUC of 0.998, indicating a well-balanced and highly effective binary classifier \ref{PERFORMANCE-CHART} \ref{table:model_metrics}. NASNetMobile and VGG16 also performed strongly, with high scores across all metrics, making them solid alternatives. ResNet50 showed competitive results but fell slightly short of the top three, particularly in precision. On the other hand, models such as EfficientNetB0, EfficientNetV2B0, ConvNeXtTiny and MobileNet performed poorly. Despite their perfect recall, their low precision and AUC values suggest they overpredicted the positive class, leading to high false positive rates. MobileNetV2, despite a decent accuracy and AUC, failed to maintain balance across precision and recall, making it less suitable for reliable classification in this context. Given its superior and consistent results, DenseNet121 stands out as the most suitable model for deployment, offering both robustness and high predictive accuracy for this binary classification task.
\begin{figure}
  \begin{center}
  \includegraphics[width=6.5in]{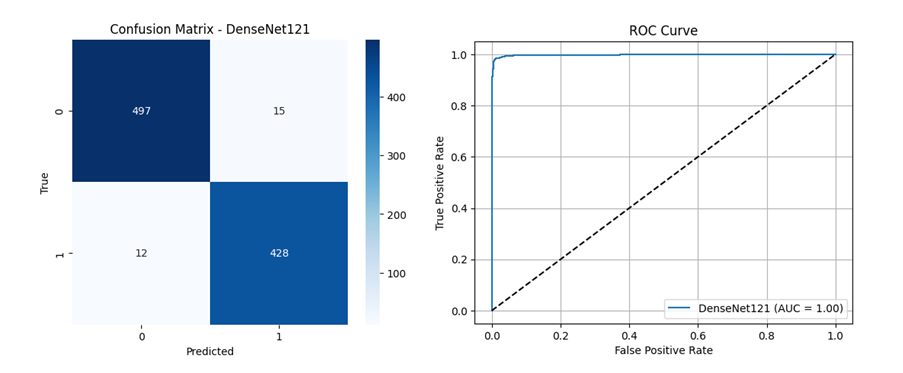}\\
  \caption{Confusion Matrix and ROC Curve for DenseNet121}\label{CONFUSION-MATRIX}
  \end{center}
\end{figure}

The confusion matrix reflects DenseNet121’s exceptional classification accuracy with minimal misclassifications:
\begin{itemize}
    \item True Negatives (TN): 497
    \item False Positives (FP): 15
    \item False Negatives (FN): 12
    \item True Positives (TP): 428
\end{itemize}

This balance indicates that the model is not only highly accurate but also well-calibrated in terms of sensitivity (recall) and specificity.

The ROC curve further supports these results, with an AUC score of 1.00, demonstrating near-perfect separation between positive and negative classes. The curve closely hugs the top-left corner, indicating excellent trade-off between true positive rate and false positive rate.

Together, these visualizations affirm DenseNet121’s reliability and robustness for the binary classification task of COVID-19 detection, outperforming other evaluated architectures in both quantitative metrics and qualitative visual assessment.

\section{Discussion}
\subsection{Summary}
The results show that DenseNet121 achieved the highest performance with 98\% accuracy, 96.8\% precision, 98.8\% recall demonstrating robust diagnostic capabilities.

\subsection{Conclusion}
This study introduces a robust deep learning framework for COVID-19 diagnosis using chest X-ray and CT imaging, emphasizing both high model performance and real-world deployment feasibility. Leveraging imaging data from 19 countries across diverse age groups, genders, and COVID-19 variants, the study employed comprehensive preprocessing, undersampling, and data augmentation techniques to ensure balanced and representative datasets.

To ensure practical deployment, models were optimized through quantization and pruning, making them lightweight and suitable for:
\begin{itemize}
    \item Web-based diagnostic platforms via cloud APIs (Flask or RESTAPI with TensorFlow Serving).
    \item Mobile applications using TensorFlow Lite or ONNX for on-device diagnosis especially valuable in low-resource and rural settings.
\end{itemize}

The framework further integrates Grad-CAM visualizations for explainability, federated learning for privacy-preserving collaboration across hospitals, and longitudinal monitoring for tracking long COVID or reinfection cases. These features collectively position the system as a clinically relevant, mutation-resilient, and scalable solution for COVID-19 screening and triage in modern healthcare environments.

For future work, we aim to extend this framework to multi-class classification, distinguishing between lung pathologies such as tuberculosis, AIDS, and COVID-19. This initiative will be pursued in collaboration with clinicians to enhance diagnostic specificity and clinical utility.

\subsection{Future Work}
\subsubsection{Clinical Validation Across Institutions}
We aim to collaborate with multiple hospitals and diagnostic centers to externally validate our model on institution-specific datasets. This will help assess the model’s generalizability and robustness across different scanners, protocols, and patient populations.

\subsection{Integration with Electronic Health Records (EHR)}
Work is underway to integrate our diagnostic tool with EHR systems for seamless access to patient history and real-time imaging data, enabling context-aware predictions and decision support.

\subsection{Deployment on Web and Mobile Platforms}
We are optimizing the final model using techniques such as quantization and pruning for deployment on edge devices and cloud platforms. This will support real-time diagnosis via a web interface and mobile app, particularly in resource-constrained or rural areas.

\subsection{Regulatory Readiness and Clinical Trials}
We are preparing documentation and performance benchmarks to pursue regulatory approval (CE marking, FDA clearance). A prospective clinical trial is also being designed to measure diagnostic impact in a real-world setting.

\subsection{Extension to Long COVID and Follow-up Monitoring}
We plan to adapt our system for longitudinal analysis, enabling clinicians to track radiological changes over time, useful for monitoring long COVID progression or reinfections.

\subsection{Federated Learning for Privacy-Preserving AI}
To support data privacy and multi-institutional collaboration, we will explore federated learning frameworks that allow model training on decentralized data without sharing patient images.

\subsection{Acknowledgements}
I would like to express my gratitude to my supervisor Li Zhang who has shaped, guided, and refined my work through this experiment. Her subject expertise, her intuition on the areas to explore and her patience as a teacher played a major part in making this project what it is today.

\subsection{Conflicts of Interest}
None declared.

\subsection{Abbreviations}
\begin{itemize}
    \item COVID-19: Coronavirus Disease 2019
    \item CNN: Convolutional Neural Network 
    \item VGG16: Visual geometry group 16-layer network
    \item RestNet50: Residual Network with 50 layers
    \item InceptionV3: Inception Version 3
    \item EfficientNetB0: Efficient Network Baseline 0
    \item Xception: Extreme Inception
    \item DenseNet121: Densely Connected Convolutional Network (121-layer)
    \item MobileNetV2: Mobile Network Version 2
    \item NASNetMobile: Neural Architecture Search Network (Mobile version)
    \item CT-scan: Computed Tomography scan
    \item X-ray: X-radiation
    \item PHEIC: Public Health Emergency of International Concern
    \item GISRS: Global Influenza Surveillance and Response System
    \item RT-PCR: Reverse transcription-quantitative polymerase chain reaction
    \item AI: Artificial Intelligence
    \item AIDS: Acquired Immunodeficiency Syndrome
    \item VOI: Variant of Interest
    \item VUM: Variants Under Monitoring
\end{itemize}

\bibliographystyle{unsrt}  
\bibliography{COVID_19_Pneumonia_Diagnosis_Using_Medical_Images_Deep_Learning_Based_Transfer_Learning_Approach}

\begin{thebibliography}{10}

\bibitem{ref1}
World~Health Organization.
\newblock Covid-19 - global situation, 2025.

\bibitem{ref2}
Ada Health.
\newblock Covid-19 variants: 2024 dominant variants and symptoms, 2025.

\bibitem{ref6}
World~Health Organization.
\newblock Pneumonia of unknown cause - china, 2020.

\bibitem{ref7}
World~Health Organization.
\newblock Coronavirus, n.d.

\bibitem{ref9}
World~Health Organization.
\newblock Weekly epidemiological update on covid-19 - 25 august 2023, 2023.

\bibitem{ref10}
World~Health Organization.
\newblock Covid-19 public health emergency of international concern (pheic) global research and innovation forum, 2020.

\bibitem{ref3}
Yale Medicine.
\newblock 3 things to know about xec, the dominant covid strain, 2024.

\bibitem{ref4}
Centers for Disease~Control and Prevention.
\newblock Covid data tracker, 2024.

\bibitem{ref5}
Oceanit.
\newblock Who’s 2025 updates on covid-19 variants: Focus on xec, testing, and recovery – assure rapid tests, 2025.

\bibitem{ref11}
World~Health Organization.
\newblock Tracking sars-cov-2 variants, 2025.

\bibitem{ref8}
World~Health Organization.
\newblock Coronavirus, n.d.

\bibitem{ref12}
Ada Health.
\newblock Omicron vs delta comparison: Risks, difference, treatments | ada health, 2025.

\bibitem{ref26}
Ullah SMA; Islam MM; Mahmud S; Nooruddin S; Raju SMTU;~Haque MR.
\newblock Scalable telehealth services to combat novel coronavirus (covid-19) pandemic, 2021.
\newblock PMID: 33426530; PMCID: PMC7786340.

\bibitem{ref27}
Zhong Qiu; Wong.~Alexander Wang, Linda;~Lin.
\newblock Covid-net: a tailored deep convolutional neural network design for detection of covid-19 cases from chest x-ray images, 2020.

\bibitem{ref30}
Reshan MSA; Gill KS; Anand V; Gupta S; Alshahrani H; Sulaiman A;~Shaikh A.
\newblock Detection of pneumonia from chest x-ray images utilizing mobilenet model, 2023.

\bibitem{ref31}
Muhammad; Furqan Rustam; Roberto Álvarez; Juan Luis Vidal Mazón; Isabel de la Torre Díez; Imran~Ashraf Mujahid.
\newblock Pneumonia classification from x-ray images with inception-v3 and convolutional neural network, 2022.

\bibitem{ref28}
Mustafa Ghaderzadeh.
\newblock Deep convolutional neural network–based computer-aided detection system for covid-19 using multiple lung scans: Design and implementation study, 2021.
\newblock PMID: 33848973; PMCID: 8078376.

\bibitem{ref29}
Jiang W; Ji W; Zhang Y; Xie Y; Chen S; Jin Y;~Duan G.
\newblock An update on detection technologies for sars-cov-2 variants of concern, 2022.
\newblock PMID: 36366421; PMCID: PMC9693800.

\bibitem{ref32}
Miró Catalina Q; Fuster-Casanovas A; Solé-Casals J; Vidal-Alaball J.
\newblock Developing an artificial intelligence model for reading chest x-rays: Protocol for a prospective validation study, 2022.
\newblock Accessed: 2025-07-07.

\bibitem{ref33}
Wang Qian ; Mellis Ian A; Ho Jonathan ; Bowen Andrew ; Kowalski-Dobson Thomas ; Valdez Raul ; Katsamba Panagiotis S. ; Wu Ming ; Lee Charles ; Shapiro Lawrence ; Gordon Ari ; Guo Yu ; Ho David D. ;~Liu Linfa.
\newblock Recurrent sars-cov-2 spike mutations confer growth advantages to select jn.1 sublineages.
\newblock {\em Emerging Microbes \& Infections}, 13(1):2402880, 2024.

\bibitem{lidc-idri}
K~Scott Mader.
\newblock The lung image database consortium image collection (lidc-idri), 2021.

\bibitem{ref24}
SIRM.
\newblock Sirm, n.d.

\bibitem{ref13}
BIMCV.
\newblock Bimcv-covid19 – bimcv, 2023.

\bibitem{ref25}
CNCB.
\newblock Ictcf, n.d.

\bibitem{ref16}
The Cancer Imaging~Archive (TCIA).
\newblock ct-images-in-covid-19, n.d.

\bibitem{ref19}
The Cancer Imaging~Archive (TCIA).
\newblock Midrc-ricord-1a (covid), n.d.

\bibitem{ref17}
The Cancer Imaging~Archive (TCIA).
\newblock Midrc-ricord-1b (normal), n.d.

\bibitem{ref18}
The Cancer Imaging~Archive (TCIA).
\newblock Midrc-ricord-1c (covid), n.d.

\bibitem{ref21}
Grand Challenge.
\newblock Stoic2021 - covid-19 ai challenge - grand challenge, n.d.

\bibitem{ref22}
Radiopaedia.
\newblock Covid-19, n.d.

\bibitem{ref23}
Diagnostics and telemedicine center.
\newblock Diagnostics and telemedicine center, n.d.

\bibitem{ref34}
Anjali Dharmik.
\newblock Covid-19-app, 2025.
\newblock Accessed: 2025-07-07.

\end{thebibliography}

\end{document}